\documentclass[12pt,english,onecolumn]{IEEEtran}
\usepackage[T1]{fontenc}
\usepackage{amsmath}
\usepackage{graphicx}
\usepackage{amssymb}

\makeatletter
 \newtheorem{thm}{Theorem}
 \newtheorem{prop}{Proposition}
 \usepackage{verbatim}

\usepackage{cite}

\usepackage{babel}
\makeatother
\begin{document}

\title{Multi-Access MIMO Systems with Finite Rate Channel State Feedback}

\author{Wei Dai, Brian Rider and Youjian Liu\\
dai@colorado.edu, brider@euclid.colorado.edu, and eugeneliu@ieee.org\\
University of Colorado at Boulder}

\maketitle
\begin{abstract}
This paper characterizes the effect of finite rate channel state feedback
on the sum rate of a multi-access multiple-input multiple-output (MIMO)
system. We propose to control the users jointly, specifically, we
first choose the users jointly and then select the corresponding beamforming
vectors jointly. To quantify the sum rate, this paper introduces the
\emph{composite Grassmann manifold} and the \emph{composite Grassmann
matrix.} By characterizing the distortion rate function on the composite
Grassmann manifold and calculating the logdet function of a random
composite Grassmann matrix, a good sum rate approximation is derived.
According to the distortion rate function on the composite Grassmann
manifold, the loss due to finite beamforming decreases exponentially
as the feedback bits on beamforming increases.
\end{abstract}
\begin{keywords}
multi-access, MIMO, limited feedback
\end{keywords}

\section{\label{sec:Introduction}Introduction}

This paper considers the uplink of a cellular system with one base
station and multiple users, where both the base station and each user
are equipped with multiple antennas. Multiple antenna systems, also
known as multiple-input multiple-output (MIMO) systems, provide significant
benefit over single antenna systems in terms of either higher spectral
efficiency or better reliability. For the uplink of a cellular system,
it is reasonable to assume that the base station has the full knowledge
about the uplink channel while the users has partial information about
the uplink channel through a feedback link from the base station.
In practice, it is also reasonable to assume that the feedback link
is rate limited.

The purpose of this paper is to quantify the effect of the finite
rate channel state feedback on the sum rate. The effect of finite
rate feedback on single user MIMO systems are well studied. MIMO systems
with only one on-beam are considered in \cite{Sabharwal_IT03_Beamforming_MIMO}
and \cite{Love_IT03_Grassman_Beamforming_MIMO} while systems with
multiple on-beams are discussed in \cite{Honig_Allerton03_Benefits_Limited_Feedback_Wireless_Channels,Rao_icc05_MIMO_spatial_multiplexing_limit_feedback,Love_IT05sub_limited_feedback_unitary_precoding,Heath_ICASSP05_Quantization_Grassmann_Manifold,Dai_ISIT05_Power_onoff_strategy_design,Dai_Globecom05_Quantization_bounds_Grassmann_manifold}.
In the recent works \cite{Dai_ISIT05_Power_onoff_strategy_design}
and \cite{Dai_Globecom05_Quantization_bounds_Grassmann_manifold},
the effect of finite rate feedback is accurately quantified by characterizing
the distortion rate function in the Grassmann manifold. For multi-access
systems, the throughput capacity region is characterized in \cite{Tse_IT98_multiaccess_I_optimal_resource_allocation}
with the assumption that each user has only one antenna and the full
channel information is available to all users. 

To characterize the feedback gain, we propose to control the users
jointly. An simple extension of \cite{Lau_IT04_Capacity_Memoryless_Block_Fading}
can show that the optimal strategy is to select the covariance matrices
of the transmit signals of the users jointly. It is different from
the current systems where the base station controls the users individually.
The gain of joint control over individual control is analogous to
that of vector quantization over scalar quantization. However, it
is difficult to either implement or analyze the fully joint control.
For simplicity, this paper proposes a suboptimal strategy employing
power on/off strategy, where we first choose the on-users jointly
and then select the beamforming vectors jointly. The effect of user
choice can be analyzed by extreme order statistics. To quantify the
effect of beamforming, the \emph{composite Grassmann manifold} is
introduced in this paper. By characterizing the distortion rate function
on the composite Grassmann manifold and calculating the logdet function
of a random \emph{composite Grassmann matrix}, a good sum rate approximation
is derived. According to the distortion rate function on the composite
Grassmann manifold, the loss of finite beamforming decreases exponentially
as the feedback bits on beamforming increases.

\section{\label{sec:System-Model}System Model}

Assume that there are $L_{R}$ antennas at the base station and $N$
users communicating with the base station. Assume that the user $i$
has $L_{T,i}$ antennas $1\leq i\leq N$. In this paper, we let $L_{T,i}=L_{T,j}=L_{T}$
for $1\leq i,j\leq N$. The signal transmission model is \[
\mathbf{Y}=\sum_{i=1}^{N}\mathbf{H}_{i}\mathbf{T}_{i}+\mathbf{W},\]
where $\mathbf{Y}\in\mathbb{C}^{L_{R}\times1}$ is the received signal
at the base station, $\mathbf{H}_{i}\in\mathbb{C}^{L_{R}\times L_{T}}$
is the channel state matrix for user $i$, $\mathbf{T}_{i}$ is the
transmitted Gaussian signal vector for user $i$ and $\mathbf{W}\in\mathbb{C}^{L_{R}\times1}$
is the additive Gaussian noise vector with zero mean and covariance
matrix $\mathbf{I}_{L_{R}}$. In this paper, we assume the Rayleigh
fading channel model, i.e., the entries of $\mathbf{H}_{i}$ are independent
and identically distributed (i.i.d.) circularly symmetric complex
Gaussian variables with zero mean and unit variance ($\mathcal{CN}\left(0,1\right)$)
and $\mathbf{H}_{i}$'s are independent for each channel use. 

We assume that there exists a common feedback link from the base station
to all the users. At the beginning of each channel use, the channel
states $\mathbf{H}_{i}$'s are perfectly estimated at the receiver.
A message, which is a function of the channel state, is sent back
to all users through a feedback channel. The feedback is error-free
and rate limited. The feedback directs the users to choose their Gaussian
signal covariance matrices. In multi-access system, users are uncoordinated.
It is reasonable to assume that $\mathrm{E}\left[\mathbf{T}_{i}\mathbf{T}_{j}^{\dagger}\right]=\mathbf{0}$.
Let $\mathbf{T}=\left[\mathbf{T}_{1}^{\dagger}\cdots\mathbf{T}_{N}^{\dagger}\right]^{\dagger}$
be the overall transmitted Gaussian signal for all users and $\mathbf{\Sigma}\triangleq\mathrm{E}\left[\mathbf{T}\mathbf{T}^{\dagger}\right]$
be the overall signal covariance matrix. Then $\mathbf{\Sigma}$ is
an $NL_{T}\times NL_{T}$ block diagonal matrix whose $i^{\mathrm{th}}$
diagonal block is the $L_{T}\times L_{T}$ covariance matrix $\mathrm{E}\left[\mathbf{T}_{i}\mathbf{T}_{i}^{\dagger}\right]$.
Assume there is a covariance matrix codebook $\mathcal{B}_{\mathbf{\Sigma}}=\left\{ \mathbf{\Sigma}_{1},\cdots,\mathbf{\Sigma}_{K_{\mathcal{B}}}\right\} $
declared to both the base station and the users, where each $\mathbf{\Sigma}_{k}$
is a proper overall signal covariance matrix and $K_{\mathcal{B}}$
is the size of the codebook. Let $\mathbf{H}=\left[\mathbf{H}_{1}\mathbf{H}_{2}\cdots\mathbf{H}_{N}\right]$
be the overall channel state matrix. The feedback function $\varphi$
is a mapping from $\left\{ \mathbf{H}\in\mathbb{L}^{L_{R}\times NL_{T}}\right\} $
into the index set $\left\{ 1,\cdots,K_{\mathcal{B}}\right\} $. Subjected
to the finite rate feedback constraint\[
\left|\mathcal{B}_{\mathbf{\Sigma}}\right|\leq K_{\mathcal{B}}\]
 and the average transmission power constraint\[
\mathrm{E}_{\mathbf{H}}\left[\mathrm{tr}\left(\mathbf{\Sigma}_{\varphi\left(\mathbf{H}\right)}\right)\right]\leq\rho,\]
we are interested in characterizing the sum rate \begin{equation}
\underset{\mathcal{B}_{\mathbf{\Sigma}}}{\max}\;\underset{\varphi\left(\cdot\right)}{\max}\;\mathrm{E}_{\mathbf{H}}\left[\log\left|\mathbf{I}_{L_{R}}+\mathbf{H}\mathbf{\Sigma}_{\varphi\left(\mathbf{H}\right)}\mathbf{H}^{\dagger}\right|\right].\label{eq:sum_rate_optimal}\end{equation}
Since the variance of the Gaussian noise is normalized, the average
power constraint $\rho$ is also the average received signal-to-noise
ratio (SNR).

\section{\label{sec:Mathematical-Preliminary-Results}Mathematical Preliminary
Results}

For compositional clarity, this section assembles the useful mathematical
results that we derive for later analysis. Due to the space limit,
\emph{we omit all the proofs}.

\subsection{\label{sub:Extreme-Chi2}Extreme Order Statistics for Chi-Square
Random Variable}

Let $X_{i}=\sum_{j=1}^{L}\left|h_{i,j}\right|^{2}$ where $h_{i,j}\;1\leq j\leq L,\;1\leq i\leq n$
are i.i.d. circularly symmetric complex Gaussian variables with zero
mean and unit variance. Let us rearrange these i.i.d. chi-square random
variables $X_{1},\cdots,X_{n}$ into a nondecreasing sequence $X_{i_{1}}\leq X_{i_{2}}\leq\cdots\leq X_{i_{n}}$.
Let $n$ approach infinity, the following theorem gives a formula
for $\mathrm{E}\left[\sum_{k=1}^{l}X_{i_{n-k+1}}\right]$ where $l$
is a fixed positive integer. 

\begin{thm}
\label{thm:Expectation-extreme-chi2}Let $X=\sum_{j=1}^{L}\left|h_{j}\right|^{2}$
where $h_{j}\sim\mathcal{CN}\left(0,1\right)$. Denote the distribution
function of $X$ by $F_{X}\left(x\right)$. Then for any fixed positive
integer $l$, \[
\underset{n\rightarrow+\infty}{\lim}\;\mathrm{E}\left[\frac{\sum_{k=1}^{l}X_{i_{n-k+1}}-la_{n}}{b_{n}}\right]=l\left(\mu_{1}^{x}+1-\sum_{k=1}^{l}\frac{1}{k}\right),\]
where $a_{n}$ is the solution of \[
a_{n}=\inf\left\{ x:\;1-F_{X}\left(x\right)\leq\frac{1}{n}\right\} ,\]
\[
b_{n}=\frac{\sum_{i=0}^{L-1}\frac{L-i}{i!}a_{n}^{i}}{\sum_{i=0}^{L-1}\frac{1}{i!}a_{n}^{i}},\]
and\[
\mu_{1}^{x}=\int_{-\infty}^{+\infty}xde^{-e^{-x}}=0.577216\cdots.\]

\end{thm}
Although this theorem is for asymptotically large $n$, it gives an
accurate approximation when $0<l\ll n$.

\subsection{\label{sub:Conditioned-Eigen}Conditioned Eigenvalues of the Wishart
Matrix}

Let $\mathbf{H}\in\mathbb{L}^{n\times m}$ be a random $n\times m$
matrix whose entries are i.i.d. Gaussian random variables with zero
mean and unit variance, where $\mathbb{L}$ is either $\mathbb{R}$
or $\mathbb{C}$ and $m\leq n$ w.l.o.g.. The random matrix $\mathbf{W}=\mathbf{H}^{\dagger}\mathbf{H}$
is Wishart distributed and its distribution is denoted by $W_{m}\left(n,\mathbf{I}_{m}\right)$.

For a $\mathbf{W}\sim W_{m}\left(n,\mathbf{I}_{m}\right)$, the following
proposition shows that conditioned on the trace, the conditional expectation
of a specific eigenvalue of $\mathbf{W}$ is proportional to the condition
with a ratio independent of that condition.

\begin{prop}
\label{pro:conditional-expectation-Wishart}Let $\mathbf{W}\sim W_{m}\left(n,\mathbf{I}_{m}\right)$
where $n\geq m$. List the ordered eigenvalues of $\mathbf{W}$ as
$\lambda_{1}\geq\lambda_{2}\geq\cdots\geq\lambda_{m}\geq0$. Then
conditioned on the trace of $\mathbf{W}$, i.e., $\sum_{i=1}^{m}\lambda_{i}=c$
where $c>0$, the ratio between the conditional expectation of $\lambda_{i}$
and the condition $c$ is a constant $\zeta_{i}$ independent of $c$,
i.e.,\[
\mathrm{E}\left[\lambda_{i}|\sum_{i=1}^{m}\lambda_{i}=c\right]=\zeta_{i}c\]
where \[
\zeta_{i}=\frac{\int_{\underset{\lambda_{1}\geq\cdots\geq\lambda_{m}}{\sum\lambda_{j}=1}}\lambda_{i}\prod_{j=1}^{m}\lambda_{i}^{\frac{\beta}{2}\left(n-m+1\right)-1}\left|\Delta_{m}\left(\mathbf{\lambda}\right)\right|^{\beta}\prod_{j=1}^{m}d\lambda_{j}}{\int_{\underset{\lambda_{1}\geq\cdots\geq\lambda_{m}}{\sum\lambda_{j}=1}}\prod_{j=1}^{m}\lambda_{i}^{\frac{\beta}{2}\left(n-m+1\right)-1}\left|\Delta_{m}\left(\mathbf{\lambda}\right)\right|^{\beta}\prod_{j=1}^{m}d\lambda_{j}},\]
$\beta=1$ if $\mathbb{L}=\mathbb{R}$ or $\beta=2$ if $\mathbb{L}=\mathbb{C}$,
and $\left|\Delta_{m}\left(\mathbf{\lambda}\right)\right|=\prod_{i<j}^{m}\left(\lambda_{i}-\lambda_{j}\right)$. 
\end{prop}

In general, it is not easy to calculate the constant $\zeta_{i}$
$1\leq i\leq m$. Fortunately, the constants can be well approximated
by asymptotics. Due to the space limit, we only present the asymptotic
formula for $\zeta_{1}$ in the following proposition.

\begin{prop}
\label{pro:conditioned-expectation-lambda1}Let the random matrix
$\mathbf{W}\sim W_{m}\left(n,\mathbf{I}_{m}\right)$ where $n\geq m$.
Define $y\triangleq\frac{m}{n}$. Then the asymptotic approximation
gives \[
\mathrm{E}\left[\lambda_{1}|\sum_{i=1}^{m}\lambda_{i}=c\right]\approx\frac{1}{\pi}\left[\pi-a+\frac{1}{2}\sin\left(2a\right)\right]c,\]
 where $a$ satisfies \[
\frac{1}{m}=\left\{ \begin{array}{ll}
\frac{1}{\pi}\left[\pi-a-\frac{1}{\sqrt{y}}\sin\left(a\right)+\frac{1-y}{y}\theta_{y}\right] & \mathrm{if}\; y<1\\
\frac{1}{\pi}\left[\pi-a-\sin\left(a\right)\right] & \mathrm{if}\; y=1\end{array}\right.,\]
and \[
\theta_{y}=\tan^{-1}\left(\frac{\sqrt{y}\sin\left(a\right)}{1-\sqrt{y}\cos\left(a\right)}\right).\]

\end{prop}

\subsection{\label{sub:CGManifold}The Grassmann Manifold and the Composite Grassmann
Manifold}

The Grassmann manifold is the geometric object relevant to the beamforming
quantization analysis. The Grassmann manifold $\mathcal{G}_{n,m}\left(\mathbb{L}\right)$
is the set of $m$-dimensional planes (passing through the origin)
in Euclidean $n$-space $\mathbb{L}^{n}$. A generator matrix $\mathbf{P}\in\mathbb{L}^{n\times m}$
for an $m$-plane $P\in\mathcal{G}_{n,m}\left(\mathbb{L}\right)$
is the matrix whose columns are orthonormal and span $P$. The generator
matrix is not unique. That is, if $\mathbf{P}$ generates $P$ then
$\mathbf{PU}$ also generates $P$ for any $m\times m$ orthogonal/unitary
matrix $\mathbf{U}$ (w.r.t. $\mathbb{L}=\mathbb{R}/\mathbb{C}$ respectively)
\cite{Conway_96_PackingLinesPlanes}. The chordal distance between
two $m$-planes $P_{1},P_{2}\in\mathcal{G}_{n,m}\left(\mathbb{L}\right)$
can be defined by their generator matrices $\mathbf{P}_{1}$ and $\mathbf{P}_{2}$
via $d_{c}\left(P_{1},P_{2}\right)$$=\frac{1}{\sqrt{2}}\left\Vert \mathbf{P}_{1}\mathbf{P}_{1}^{\dagger}-\mathbf{P}_{2}\mathbf{P}_{2}^{\dagger}\right\Vert _{F}$\cite{Conway_96_PackingLinesPlanes}.
The uniform distribution on $\mathcal{G}_{n,m}\left(\mathbb{L}\right)$
with density function $f_{P}\left(\cdot\right)$ satisfies $f_{P}\left(P_{1}\right)=f_{P}\left(P_{2}\right)$
for arbitrary $P_{1},P_{2}\in\mathcal{G}_{n,m}\left(\mathbb{L}\right)$
\cite{Muirhead_book82_multivariate_statistics}.

For quantizations on $\mathcal{G}_{n,m}\left(\mathbb{L}\right)$,
the corresponding distortion rate function has been characterized
\cite{Dai_ISIT05_Power_onoff_strategy_design}. A quantization $q$
on $\mathcal{G}_{n,m}\left(\mathbb{L}\right)$ is a mapping from $\mathcal{G}_{n,p}\left(\mathbb{L}\right)$
to a subset of $\mathcal{G}_{n,p}\left(\mathbb{L}\right)$, which
is typically called a code $\mathcal{C}$, i.e., $q:\mathcal{G}_{n,p}\left(\mathbb{L}\right)\rightarrow\mathcal{C}.$
Define the distortion metric as the squared chordal distance. Then
the distortion associated with a quantization $q$ is \[
D\triangleq\mathrm{E}_{Q}\left[d_{c}^{2}\left(Q,q\left(Q\right)\right)\right],\]
where the source $Q$ is randomly distributed in $\mathcal{G}_{n,m}\left(\mathbb{L}\right)$.
Assume that the source $Q$ is uniformly distributed in $\mathcal{G}_{n,p}\left(\mathbb{L}\right)$.
For any given code $\mathcal{C}$, the optimal quantization to minimize
the distortion is%
\footnote{The ties, i.e., the case that $\exists P_{1},P_{2}\in\mathcal{C}$
such that $d_{c}\left(P_{1},Q\right)=\underset{P\in\mathcal{C}}{\min}\; d_{c}\left(P,Q\right)=d_{c}\left(P_{2},Q\right)$,
are broken arbitrarily because the probability of ties is zero.%
} \[
q\left(Q\right)=\arg\;\underset{P\in\mathcal{C}}{\min}\; d_{c}\left(P,Q\right).\]
The distortion associated with this quantization is\[
D\left(\mathcal{C}\right)=\mathrm{E}_{Q}\left[\underset{P\in\mathcal{C}}{\min}\; d_{c}^{2}\left(P,Q\right)\right].\]
For a given code size $K$ where $K$ is a positive integer, the distortion
rate function is%
\footnote{The standard definition of the distortion rate function is a function
of the code rate defined by $\log_{2}K$. The definition in this paper
is equivalent to the standard one.%
} \[
D^{*}\left(K\right)=\underset{\mathcal{C}:\left|\mathcal{C}\right|=K}{\inf}\; D\left(\mathcal{C}\right).\]
In \cite{Dai_Globecom05_Quantization_bounds_Grassmann_manifold},
we derive a lower bound and an upper bound for $\mathbb{L}=\mathbb{C}$\[
\frac{t}{t+1}\eta^{-\frac{1}{t}}2^{-\frac{\log_{2}K}{t}}\lesssim D^{*}\left(K\right)\lesssim\frac{\Gamma\left(\frac{1}{t}\right)}{t}\eta^{-\frac{1}{t}}2^{-\frac{\log_{2}K}{t}},\]
where $t=m\left(n-m\right)$, \[
\eta=\left\{ \begin{array}{ll}
\frac{1}{t!}\prod_{i=1}^{m}\frac{\left(n-i\right)!}{\left(m-i\right)!} & \mathrm{if}\;1\leq m\leq\frac{n}{2}\\
\frac{1}{t!}\prod_{i=1}^{n-m}\frac{\left(n-i\right)!}{\left(n-m-i\right)!} & \mathrm{if}\;\frac{n}{2}\leq m\leq n\end{array}\right.,\]
and the symbol $\lesssim$ denotes the \emph{main order inequality},
$f\left(K\right)\lesssim g\left(K\right)$ if $\underset{K\rightarrow+\infty}{\lim}\frac{f\left(K\right)}{g\left(K\right)}\leq1.$

To treat multi-access MIMO systems, we define the \emph{composite
Grassmann manifold}. The $k$-composite Grassmann manifold $\mathcal{G}_{n,m}^{\left(k\right)}\left(\mathbb{L}\right)$
is a Cartesian product of $k$ $\mathcal{G}_{n,m}\left(\mathbb{L}\right)$'s.
Denote $P^{\left(k\right)}$ an element in $\mathcal{G}_{n,m}^{\left(k\right)}\left(\mathbb{L}\right)$.
\[
P^{\left(k\right)}=\left(P_{1},\cdots,P_{k}\right)\]
where $P_{i}\in\mathcal{G}_{n,m}\left(\mathbb{L}\right)$ $1\leq i\leq k$.
For any $P_{1}^{\left(k\right)},P_{2}^{\left(k\right)}\in\mathcal{G}_{n,m}^{\left(k\right)}\left(\mathbb{L}\right)$,
we define the chordal distance between them\[
d_{c}\left(P_{1}^{\left(k\right)},P_{2}^{\left(k\right)}\right)=\sqrt{\sum_{i=1}^{k}d_{c}^{2}\left(P_{1,i},P_{2,i}\right)},\]
where $P_{1}^{\left(k\right)}=\left(P_{1,1},\cdots,P_{1,k}\right)$
and $P_{2}^{\left(k\right)}=\left(P_{2,1},\cdots,P_{2,k}\right)$.
It is easy to verify that the chordal distance on $\mathcal{G}_{n,m}^{\left(k\right)}\left(\mathbb{L}\right)$
is well defined.

This paper characterizes the distortion rate function for quantizations
on $\mathcal{G}_{n,m}^{\left(k\right)}\left(\mathbb{L}\right)$. Define
the distortion metric on $\mathcal{G}_{n,m}^{\left(k\right)}\left(\mathbb{L}\right)$
as the square chordal distance on it. Assume a uniformly distributed
source $Q^{\left(k\right)}$ in $\mathcal{G}_{n,m}^{\left(k\right)}\left(\mathbb{L}\right)$.
The following theorem characterizes the distortion rate function for
quantizations on $\mathcal{G}_{n,m}^{\left(k\right)}\left(\mathbb{C}\right)$.

\begin{thm}
\label{thm:DRF-composite-Grassmann}The distortion rate function on
$\mathcal{G}_{n,m}^{\left(k\right)}\left(\mathbb{C}\right)$ is upper
bounded and lower bounded by \[
\frac{kt}{kt+1}\left(\frac{\Gamma^{k}\left(t+1\right)}{\Gamma\left(kt+1\right)}\eta^{k}\right)^{-\frac{1}{kt}}2^{-\frac{\log_{2}K}{kt}}\lesssim D^{*}\left(K\right)\lesssim\frac{\Gamma\left(\frac{1}{kt}\right)}{kt}\left(\frac{\Gamma^{k}\left(t+1\right)}{\Gamma\left(kt+1\right)}\eta^{k}\right)^{-\frac{1}{kt}}2^{-\frac{\log_{2}K}{kt}},\]
where $t=m\left(n-m\right)$, \[
\eta=\left\{ \begin{array}{ll}
\frac{1}{t!}\prod_{i=1}^{m}\frac{\left(n-i\right)!}{\left(m-i\right)!} & \mathrm{if}\;1\leq m\leq\frac{n}{2}\\
\frac{1}{t!}\prod_{i=1}^{n-m}\frac{\left(n-i\right)!}{\left(n-m-i\right)!} & \mathrm{if}\;\frac{n}{2}\leq m\leq n\end{array}\right.,\]
and the symbol $\lesssim$ denotes the \emph{main order inequality},
$f\left(K\right)\lesssim g\left(K\right)$ if $\underset{K\rightarrow+\infty}{\lim}\frac{f\left(K\right)}{g\left(K\right)}\leq1.$
\end{thm}

It is noteworthy that the upper bound is derived by computing the
average distortion over the ensemble of random codes. In practice,
we often use the upper bound as an approximation to the actual distortion
rate function.

\subsection{\label{sub:CGMatrix}Composite Grassmann Matrix}

Roughly speaking, a \emph{composite Grassmann matrix} is the generator
matrix for an element in $\mathcal{G}_{n,m}^{\left(k\right)}\left(\mathbb{L}\right)$.
Let $P^{\left(k\right)}=\left(P_{1},\cdots,P_{k}\right)\in\mathcal{G}_{n,m}^{\left(k\right)}\left(\mathbb{L}\right)$.
The composite matrix $\mathbf{P}^{\left(k\right)}$ generating $P^{\left(k\right)}$
is $\mathbf{P}^{\left(k\right)}=\left[\mathbf{P}_{1}\cdots\mathbf{P}_{k}\right]$
where $\mathbf{P}_{1},\cdots,\mathbf{P}_{k}$ are the generator matrices
for $P_{1},\cdots,P_{k}$ respectively. Since the generator matrix
for a plane in the Grassmann manifold is not unique, the composite
Grassmann matrix generating $P^{\left(k\right)}$ is not unique either.
Let $\mathbf{P}^{\left(k\right)}$ be a generator matrix for $P^{\left(k\right)}$.
The matrix $\mathbf{P}^{\left(k\right)}\mathbf{U}^{\left(k\right)}$,
where $\mathbf{U}^{\left(k\right)}$ is the arbitrary $km\times km$
block diagonal matrix whose $k$ diagonal blocks are $m\times m$
orthogonal/unitary matrices (w.r.t. $\mathbb{L}=\mathbb{R}/\mathbb{C}$
respectively), also generates $P^{\left(k\right)}$. In this paper,
the set of composite Grassmann matrices for $\mathcal{G}_{n,m}^{\left(k\right)}\left(\mathbb{L}\right)$
is denoted by $\mathcal{M}_{n,m}^{\left(k\right)}\left(\mathbb{L}\right)$.

For a random composite Grassmann matrix $\mathbf{P}^{\left(k\right)}$,
the following theorem bounds $\mathrm{E}\left[\log\left|\mathbf{I}+c\mathbf{P}^{\left(k\right)\dagger}\mathbf{P}^{\left(k\right)}\right|\right]$.

\begin{thm}
\label{thm:bds_CGMatrix}Let $\mathbf{P}^{\left(k\right)}\in\mathcal{M}_{n,1}^{\left(k\right)}\left(\mathbb{L}\right)$
be uniformly distributed. For any positive constant $c$,\begin{eqnarray*}
\mathrm{E}_{\mathbf{H}}\left[\log\left|\mathbf{I}_{k}+\frac{c}{n}\mathbf{H}^{\dagger}\mathbf{H}\right|\right] & \leq & \mathrm{E}_{\mathbf{P}^{k}}\left[\log\left|\mathbf{I}_{k}+c\mathbf{P}^{\left(k\right)\dagger}\mathbf{P}^{\left(k\right)}\right|\right]\\
 & \leq & \log\mathrm{E}_{\mathbf{P}^{k}}\left[\left|\mathbf{I}_{k}+c\mathbf{P}^{\left(k\right)\dagger}\mathbf{P}^{\left(k\right)}\right|\right],\end{eqnarray*}
where $\mathbf{H}\in\mathbb{L}^{n\times k}$ has i.i.d. Gaussian entries
with zero mean and unit variance. 
\end{thm}
In the above theorem, both bounds can be computed explicitly. In \cite{Dai_05_Power_onoff_strategy_design_finite_rate_feedback},
we derive an asymptotic formula to approximate the lower bound. Let
$n$ and $k$ approach infinity simultaneously with fixed ratio, 

\begin{eqnarray*}
 &  & \underset{\left(n,k\right)\rightarrow+\infty}{\lim}\frac{1}{\min\left(n,k\right)}\mathrm{E}_{\mathbf{H}}\left[\log\left|\mathbf{I}_{k}+\frac{c}{n}\mathbf{H}^{\dagger}\mathbf{H}\right|\right]\\
 &  & =\log\left(w\right)-\log\left(\alpha\right)-\frac{u}{r}-\frac{\left(1-y\right)\log\left(1-ur\right)}{y},\end{eqnarray*}
where $y\triangleq\frac{\min\left(n,k\right)}{\max\left(n,k\right)}$,
$r\triangleq\sqrt{y}$, $\alpha\triangleq\frac{n}{\min\left(n,k\right)\cdot c}$,
$w\triangleq\frac{1}{2}\left(1+y+\alpha+\sqrt{\left(1+y+\alpha\right)^{2}-4y}\right)$
and $u\triangleq\frac{1}{2r}\left(1+y+\alpha-\sqrt{\left(1+y+\alpha\right)^{2}-4y}\right)$.
Formulas for the upper bound are also derived in this paper. Due to
the space limit, we only present the formulas for $1\leq k\leq5$.
The expectation $\mathrm{E}_{\mathbf{P}^{k}}\left[\left|\mathbf{I}_{k}+c\mathbf{P}^{\left(k\right)\dagger}\mathbf{P}^{\left(k\right)}\right|\right]$
can be calculated by 

\begin{description}
\item [k=1]$1+c$;
\item [k=2]$\left(1+c\right)^{2}-c^{2}\frac{1}{n};$
\item [k=3]$\left(1+c\right)^{3}-c^{2}\left(1+c\right)\frac{3}{n}+c^{3}\frac{2}{n^{2}};$
\item [k=4]$\left(1+c\right)^{4}-c^{2}\left(1+c\right)^{2}\frac{6}{n}+c^{3}\left(1+c\right)\frac{8}{n^{2}}-c^{4}\left(\frac{6}{n^{3}}-\frac{3}{n^{2}}\right);$
and
\item [k=5]$\left(1+c\right)^{5}-c^{2}\left(1+c\right)^{3}\frac{10}{n}+c^{3}\left(1+c\right)^{2}\frac{20}{n^{2}}-c^{4}\left(1+c\right)\left(\frac{30}{n^{3}}-\frac{15}{n^{2}}\right)+c^{5}\left(\frac{24}{n^{4}}-\frac{20}{n^{3}}\right).$
\end{description}

\section{\label{sec:Suboptimal-Feedback-Strategies}The Suboptimal Strategy
and the Sum Rate}

This section is devoted to calculate the sum rate of a multi-access
MIMO system with finite rate feedback. The computation of the sum
rate (\ref{eq:sum_rate_optimal}) involves two correlated optimization
problems: one is with respect to the feedback function $\varphi$
and the other optimization is over all possible covariance matrix
codebooks. The direct calculation of (\ref{eq:sum_rate_optimal})
is difficult. 

To reduce the complexity, we propose a suboptimal strategy to control
the users jointly. Specifically, we first choose the on-users jointly
and then select the corresponding beamforming vectors jointly. It
is different from the current system where users are controlled individually.

The assumptions for transmission are as follows. 

\begin{description}
\item [T1)]Power on/off strategy. In power on/off strategy, the user $i$'s
covariance matrix is of the form $\mathbf{\Sigma}_{i}=P_{\mathrm{on}}\mathbf{Q}_{i}\mathbf{Q}_{i}^{\dagger}$,
where $P_{\mathrm{on}}$ is a fixed positive constant to denote on-power
and $\mathbf{Q}_{i}$ is the beamforming matrix for user $i$. Denote
each column of $\mathbf{Q}_{i}$ an \emph{on-beam} and the number
of the columns of $\mathbf{Q}_{i}$ by $l_{i}$, then $\mathbf{Q}_{i}^{\dagger}\mathbf{Q}_{i}=\mathbf{I}_{l_{i}}$
where $0\leq l_{i}\leq L_{T}$ and $l_{i}=0$ is for the case that
the user $i$ is off. This assumption is motivated by the fact that
power on/off strategy is near-optimal for single user MIMO systems
\cite{Dai_ISIT05_Power_onoff_strategy_design}.
\item [T2)]At most one on-beam per user. This assumption implies either
$l_{i}=0$ or $l_{i}=1$. It is proposed so that each user has larger
probability to be turned on.
\item [T3)]Constant number of on-beams for a given SNR. Let $l=\sum_{i=1}^{N}l_{i}$
be the total number of on-beams. we assume that $l$ is a constant
independent of the specific channel realization for a given SNR. This
assumption is motivated by the fact that constant number of on-beams
is near optimal for single user systems \cite{Dai_ISIT05_Power_onoff_strategy_design}.
It will be validated for multi-access systems in later analysis.
\end{description}

The feedback is described as below.

\begin{description}
\item [F1)]User selection criterion. Assume that $l$ users will be turned
on. We choose the $l$ users with the largest channel state Frobenius
norms, i.e. $\left\Vert \mathbf{H}_{i_{j}}\right\Vert \geq\left\Vert \mathbf{H}_{i}\right\Vert $
for all $i\notin\left\{ i_{j}:\;1\leq j\leq l\right\} $ where $\left\Vert \cdot\right\Vert $
is the Frobenius norm and $i_{1},\cdots,i_{l}$ are the users chosen
to be on (\emph{on-users}).
\end{description}
According to this user selection criterion, the feedback contains
two parts, one of which indicates the $l$ on-users and the other
of which is for beamforming. Let $i_{1},\cdots,i_{l}$ be the on-users
and $\mathbf{b}_{1},\cdots,\mathbf{b}_{l}$ be the beamforming vectors
for those users. Then $\mathbf{B}=\left[\mathbf{b}_{1}\cdots\mathbf{b}_{l}\right]\in\mathcal{M}_{L_{T},1}^{\left(l\right)}\left(\mathbb{C}\right)$
where $\mathcal{M}_{L_{T},1}^{\left(l\right)}\left(\mathbb{C}\right)$
is the set of composite Grassmann matrix (Section \ref{sub:CGMatrix}).
Denote the beamforming codebook $\mathcal{B}=\left\{ \mathbf{B}_{k}:\;\mathbf{B}_{k}\in\mathcal{M}_{L_{T},1}^{\left(l\right)}\left(\mathbb{C}\right),1\leq k\leq\left|\mathcal{B}\right|\right\} $.
Then the overall feedback codebook is the Cartesian product of the
set of on-users $\left\{ \left(i_{1},\cdots,i_{l}\right)\right\} $
and the beamforming codebook $\mathcal{B}$. Let $\mathbf{H}_{i_{j}}$
be the channel state matrix for the user $i_{j}$ and $\mathbf{v}_{j,1}$
be the right singular vector corresponding to the largest singular
value of $\mathbf{H}_{i_{j}}$. Define $\mathbf{V}\triangleq\left[\mathbf{v}_{1,1}\cdots\mathbf{v}_{l,1}\right]$.
The beamforming feedback function is defined as the following.

\begin{description}
\item [F2)]Beamforming feedback function. \begin{eqnarray}
\varphi\left(\left[\mathbf{H}_{i_{1}}\cdots\mathbf{H}_{i_{l}}\right]\right) & \triangleq & \underset{1\leq k\leq\left|\mathcal{B}\right|}{\arg\;\min}\; d_{c}^{2}\left(\mathbf{V},\mathbf{B}_{k}\right)\nonumber \\
 & = & \underset{1\leq k\leq\left|\mathcal{B}\right|}{\arg\;\max}\;\sum_{j=1}^{l}\left|\mathbf{v}_{j,1}^{\dagger}\mathbf{b}_{k,j}\right|^{2},\label{eq:beam_feedback_fn}\end{eqnarray}
where $d_{c}^{2}\left(\mathbf{V},\mathbf{B}_{k}\right)$ denotes the
chordal distance between the elements in the composite Grassmann manifold
$\mathcal{G}_{L_{T},1}^{\left(l\right)}\left(\mathbb{C}\right)$ generated
by $\mathbf{V}$ and $\mathbf{B}_{k}$, and $\mathbf{b}_{k,j}$ is
the $j^{\mathrm{th}}$ column of the $k^{\mathrm{th}}$ beamforming
matrix $\mathbf{B}_{k}\in\mathcal{B}$. 
\end{description}
The feedback assumptions F1 and F2 will be validated in the later
analysis.

The above assumptions define a suboptimal strategy for multi-access
MIMO systems. The key point is that the user choice is independent
of the channel directions and the beamforming is independent of the
channel strengths (norms). In this way, the effect of user choice
and beamforming can be studied separately. Before diving into the
general analysis, we discuss a special case, antenna selection, to
get some intuition.

\subsection{Antenna Selection}

The system model for antenna selection is \[
\mathbf{Y}=\sum_{i=1}^{NL_{T}}\mathbf{h}_{i}T_{i}+\mathbf{W},\]
where $\mathbf{h}_{i}$ is the $i^{\mathrm{th}}$ column of the overall
channel state matrix $\mathbf{H}$. For each channel realization $\mathbf{H}$,
we simply choose $l$ antennas $i_{1},\cdots,i_{l}$ such that $\left\Vert \mathbf{h}_{i_{j}}\right\Vert \geq\left\Vert \mathbf{h}_{i}\right\Vert $
for all $i\notin\left\{ i_{j}:\;1\leq j\leq l\right\} $. Here, we
actually do not require one on-beam per on-user (Assumption T2). Write
$\mathbf{h}_{i_{j}}=n_{j}\mathbf{\xi}_{j}$ where $n_{j}$ is the
Frobenius norm of $\mathbf{h}_{i_{j}}$ and $\mathbf{\xi}_{j}$ is
the unit vector to present the direction of $\mathbf{h}_{i_{j}}$.
Define $\mathbf{\Xi}\triangleq\left[\mathbf{\xi}_{1}\cdots\mathbf{\xi}_{l}\right]$.
We have the following upper bound on the sum rate. \begin{eqnarray}
 &  & \mathrm{E}_{\mathbf{H}}\left[\log\left|\mathbf{I}_{L_{R}}+\frac{\rho}{l}\sum_{j=1}^{l}\mathbf{h}_{i_{j}}\mathbf{h}_{i_{j}}^{\dagger}\right|\right]\nonumber \\
 &  & =\mathrm{E}_{\mathbf{H}}\left[\log\left|\mathbf{I}_{l}+\frac{\rho}{l}\mathrm{diag}\left[n_{1}^{2},\cdots,n_{l}^{2}\right]\mathbf{\Xi}^{\dagger}\mathbf{\Xi}\right|\right]\nonumber \\
 &  & \leq\mathrm{E}_{\mathbf{\Xi}}\left[\log\left|\mathbf{I}_{l}+\frac{\rho}{l}\mathrm{E}_{\mathbf{n}^{2}}\left[\mathrm{diag}\left[n_{1}^{2},\cdots,n_{l}^{2}\right]\right]\mathbf{\Xi}^{\dagger}\mathbf{\Xi}\right|\right]\nonumber \\
 &  & =\mathrm{E}_{\mathbf{\Xi}}\left[\log\left|\mathbf{I}_{l}+\frac{\rho}{l}\frac{\mathrm{E}_{\mathbf{n}^{2}}\left[\sum_{j=1}^{l}n_{j}^{2}\right]}{l}\mathbf{\Xi}^{\dagger}\mathbf{\Xi}\right|\right],\label{eq:ub_antennaSel}\end{eqnarray}
where the inequality follows from the concavity of $\log\left|\cdot\right|$
function and the fact that $n_{j}^{2}$'s and $\mathbf{\Xi}$ are
independent. Noting that $\left\Vert \mathbf{h}_{i}\right\Vert ^{2}$'s
are i.i.d. chi-square random variables, an accurate approximation
to $\mathrm{E}_{\mathbf{n}^{2}}\left[\sum_{j=1}^{l}n_{j}^{2}\right]$
can be obtained for $l\ll N$ by applying the asymptotic extreme order
statistics in Theorem \ref{thm:Expectation-extreme-chi2}. On the
other hand, it can be proved that $\mathbf{\xi}_{j}$'s are independent
and uniformly distributed unit vectors. Regarding $\mathbf{\Xi}$
as a Cartesian product of $\mathbf{\xi}_{j}$'s, $\mathbf{\Xi}$ is
also uniformly distributed in $\mathcal{M}_{L_{R},1}^{\left(l\right)}$,
the set of composite Grassmann matrix. According to the results in
Section \ref{sub:CGMatrix} for $\mathrm{E}\left[\log\left|\mathbf{I}+c\mathbf{\Xi}^{\dagger}\mathbf{\Xi}\right|\right]$,
the upper bound of the sum rate (\ref{eq:ub_antennaSel}) can be characterized.
Simulation show that the upper bound (\ref{eq:ub_antennaSel}) is
tight. The sum rate of antenna selection is then approximately characterized.

\subsection{General Beamforming}

With the assumptions T1-3, F1 and F2, the signal model for the general
beamforming is \[
\mathbf{Y}=\sum_{j=1}^{l}\mathbf{H}_{i_{j}}\mathbf{b}_{\varphi\left(\mathbf{H}\right),j}T_{j}+\mathbf{W},\]
where $\mathbf{b}_{\varphi\left(\mathbf{H}\right),j}$ is the $j^{\mathrm{th}}$
column of the feedback beamforming matrix $\mathbf{B}_{\varphi\left(\mathbf{H}\right)}\in\mathcal{B}$.
For notational convenience, we denote $\mathbf{b}_{\varphi\left(\mathbf{H}\right),j}$
by $\mathbf{b}_{j}^{*}$ and the equivalent channel vector $\mathbf{H}_{i_{j}}\mathbf{b}_{j}^{*}$
by $\hat{\mathbf{h}}_{j}$. Let $n_{j}$ be the Frobenius norm of
$\hat{\mathbf{h}}_{j}$, $\mathbf{\xi}_{j}$ be the unit vector presenting
the direction of $\hat{\mathbf{h}}_{j}$ and $\mathbf{\Xi}=\left[\mathbf{\xi}_{1}\cdots\mathbf{\xi}_{l}\right]$.
Then the sum rate is given by\begin{eqnarray*}
 &  & \mathrm{E}_{\mathbf{H}}\left[\log\left|\mathbf{I}_{L_{R}}+\frac{\rho}{l}\sum_{j=1}^{l}\hat{\mathbf{h}}_{j}\hat{\mathbf{h}}_{j}^{\dagger}\right|\right]\\
 &  & =\mathrm{E}_{\mathbf{H}}\left[\log\left|\mathbf{I}_{l}+\frac{\rho}{l}\mathrm{diag}\left[n_{1},\cdots,n_{l}^{2}\right]\mathbf{\Xi}^{\dagger}\mathbf{\Xi}\right|\right].\end{eqnarray*}

It can be proved that $\mathbf{\xi}_{j}$'s are uniformly distributed
and independent of $n_{j}$'s. Denote the singular value decomposition
of $\mathbf{H}_{i_{j}}$ by $\mathbf{U}_{j}\mathbf{\Lambda}_{j}\mathbf{V}_{j}^{\dagger}$.
After beamforming, the equivalent channel vector for user $i_{j}$
is $\hat{\mathbf{h}}_{j}=\mathbf{U}_{j}\left(\mathbf{\Lambda}_{j}\mathbf{V}_{j}^{\dagger}\mathbf{b}_{j}^{*}\right)=\mathbf{U}_{j}\tilde{\mathbf{\xi}}_{j}n_{j}$,
where $\tilde{\mathbf{\xi}}_{j}$ is the direction of the vector $\mathbf{\Lambda}_{j}\mathbf{V}_{j}^{\dagger}\mathbf{b}_{j}^{*}$.
Since the user choice is only dependent on $\mathbf{\Lambda}_{j}$'s
and the beamforming matrix selection is only relevant to $\mathbf{V}_{j}$'s,
$\mathbf{U}_{j}$'s are independent and uniformly distributed. According
to \cite[Thm. 6.1]{James_54_Normal_Multivariate_Analysis_Orthogonal_Group},
$\mathbf{\xi}_{j}=\mathbf{U}_{j}\tilde{\mathbf{\xi}}_{j}$ is uniformly
distributed and independent of $n_{j}$'s. Thus, similar to (\ref{eq:ub_antennaSel}),
the sum rate of general beamforming can be upper bounded by \begin{equation}
\mathrm{E}_{\mathbf{\Xi}}\left[\log\left|\mathbf{I}_{l}+\frac{\rho}{l}\frac{\mathrm{E}_{\mathbf{n}^{2}}\left[\sum_{j=1}^{l}n_{j}^{2}\right]}{l}\mathbf{\Xi}^{\dagger}\mathbf{\Xi}\right|\right],\label{eq:ub_beamforming}\end{equation}
where $\mathbf{\Xi}=\left[\mathbf{\xi}_{1}\cdots\mathbf{\xi}_{l}\right]$.

It is more involved to calculate $\mathrm{E}\left[\sum_{j=1}^{l}n_{j}^{2}\right]$.
Let $\lambda_{j,k}$ $1\leq k\leq L_{T}$ be the ordered eigenvalues
of $\mathbf{H}_{i_{j}}^{\dagger}\mathbf{H}_{i_{j}}$ such that $\lambda_{j,1}\geq\lambda_{j,2}\geq\cdots\geq\lambda_{j,L_{T}}\geq0$.
Let $\mathbf{v}_{j,k}$ be the right singular vector of $\mathbf{H}_{i_{j}}$
corresponding to the $k^{\mathrm{th}}$ largest singular value $\sqrt{\lambda_{j,k}}$.
Then \begin{eqnarray}
\mathrm{E}\left[n_{j}^{2}\right] & = & \mathrm{E}\left[\left\Vert \mathbf{H}_{i_{j}}\mathbf{b}_{j}^{*}\right\Vert ^{2}\right]=\mathrm{E}\left[\mathbf{b}_{j}^{*\dagger}\mathbf{H}_{i_{j}}^{\dagger}\mathbf{H}_{i_{j}}\mathbf{b}_{j}^{*}\right]\nonumber \\
 & = & \mathrm{E}\left[\sum_{k=1}^{L_{T}}\lambda_{j,k}\left|\mathbf{v}_{j,k}^{\dagger}\mathbf{b}_{j}^{*}\right|^{2}\right]\nonumber \\
 & = & \sum_{k=1}^{L_{T}}\mathrm{E}\left[\lambda_{j,k}\right]\mathrm{E}\left[\left|\mathbf{v}_{j,k}^{\dagger}\mathbf{b}_{j}^{*}\right|^{2}\right].\label{eq:E-n^2-seperation}\end{eqnarray}
where the last equality follows from the fact that the beamforming
is independent of the channel norms, i.e., $\sum_{k=1}^{L_{T}}\lambda_{j,k}$'s.
The $\mathrm{E}\left[\lambda_{j,k}\right]$'s can be calculated by\begin{equation}
\mathrm{E}\left[\lambda_{j,k}\right]=\mathrm{E}\left[\mathrm{E}\left[\lambda_{j,k}\left|\left\Vert \mathbf{H}_{i_{j}}\right\Vert ^{2}\right.\right]\right]=\zeta_{k}\mathrm{E}\left[\left\Vert \mathbf{H}_{i_{j}}\right\Vert ^{2}\right],\label{eq:cal-lambda-i-j}\end{equation}
where the last equality is a direct application of Proposition \ref{pro:conditional-expectation-Wishart}
in Section \ref{sub:Conditioned-Eigen}. To evaluate $\mathrm{E}\left[\left|\mathbf{v}_{j,k}^{\dagger}\mathbf{b}_{j}^{*}\right|^{2}\right]$,
we need the following proposition.

\begin{prop}
\label{pro:corr_beams}Consider the beamforming feedback function
in (\ref{eq:beam_feedback_fn}). Define $\gamma\triangleq\mathrm{E}\left[\sum_{j=1}^{l}\left|\mathbf{v}_{j,1}^{\dagger}\mathbf{b}_{j}^{*}\right|^{2}\right]$.
Then $\mathrm{E}\left[\left|\mathbf{v}_{j,1}^{\dagger}\mathbf{b}_{j}^{*}\right|^{2}\right]=\frac{\gamma}{l}$
and $\mathrm{E}\left[\left|\mathbf{v}_{j,k}^{\dagger}\mathbf{b}_{j}^{*}\right|^{2}\right]=\left(1-\frac{\gamma}{l}\right)/\left(L_{T}-1\right)$
for all $1\leq j\leq l$ and $2\leq k\leq L_{T}$.
\end{prop}
Apply this proposition and substitute (\ref{eq:cal-lambda-i-j}) into
(\ref{eq:E-n^2-seperation}). After some elementary manipulations,
we have \begin{eqnarray}
\mathrm{E}\left[\sum_{j=1}^{l}n_{j}^{2}\right] & = & \left(\frac{\zeta_{1}\gamma}{l}+\frac{\left(1-\zeta_{1}\right)\left(l-\gamma\right)}{l\left(L_{T}-1\right)}\right)\sum_{j=1}^{l}\mathrm{E}\left[\left\Vert \mathbf{H}_{i_{j}}\right\Vert ^{2}\right].\label{eq:E-n^2-calculation}\end{eqnarray}
Theorem \ref{thm:Expectation-extreme-chi2} and Proposition \ref{pro:conditioned-expectation-lambda1}
provide asymptotic formulas to approximate $\sum_{j=1}^{l}\mathrm{E}\left[\left\Vert \mathbf{H}_{i_{j}}\right\Vert ^{2}\right]$
and $\zeta_{1}$ respectively. Define $K\triangleq\left|\mathcal{B}\right|$
the size of the beamforming codebook. The maximum $\gamma$ achievable
$\gamma_{\sup}$ is a function of $K$. According to the distortion
rate function $D^{*}\left(K\right)$ for quantizations on the composite
Grassmann manifold $\mathcal{G}_{L_{T},1}^{\left(l\right)}\left(\mathbb{C}\right)$,
\[
\gamma_{\sup}\triangleq\underset{\mathcal{B}:\;\left|\mathcal{B}\right|\leq K}{\sup}\;\gamma=l-D^{*}\left(K\right).\]
Substitute $\gamma_{\sup}$ into (\ref{eq:E-n^2-calculation}). The
expectation $\mathrm{E}\left[\sum_{j=1}^{l}n_{j}^{2}\right]$ can
be calculated as a function of $K$.

Finally, substituting the value of $\mathrm{E}\left[\sum_{j=1}^{l}n_{j}^{2}\right]$
into (\ref{eq:ub_beamforming}) and employing the results in Section
\ref{sub:CGMatrix} for $\mathrm{E}\left[\log\left|\mathbf{I}+c\mathbf{\Xi}^{\dagger}\mathbf{\Xi}\right|\right]$,
the upper bound of the sum rate (\ref{eq:ub_beamforming}) can be
characterized. Simulations show that this upper bound is tight. The
sum rate is therefore approximately characterized.

\subsection{The Effect of Finite Rate Feedback}

The above analysis characterizes the effect of finite rate channel
state feedback. The upper bound (\ref{eq:ub_beamforming}) shows that
the effect of feedback is quantified by $\mathrm{E}\left[\sum_{j=1}^{l}n_{j}^{2}\right]$.
Formula (\ref{eq:E-n^2-seperation}) shows that the effect of user
choice and beamforming can be analyzed separately.

According to (\ref{eq:E-n^2-calculation}), the effects of user choice
is reflected by $\sum_{j=1}^{l}\mathrm{E}\left[\left\Vert \mathbf{H}_{i_{j}}\right\Vert ^{2}\right]$.
Maximization of the sum rate requires to maximize $\sum_{j=1}^{l}\mathrm{E}\left[\left\Vert \mathbf{H}_{i_{j}}\right\Vert ^{2}\right]$
and thus the user selection criterion (Assumption F1) is validated.
Furthermore, the term $\sum_{j=1}^{l}\mathrm{E}\left[\left\Vert \mathbf{H}_{i_{j}}\right\Vert ^{2}\right]$
is an increasing function of the number of users $N$ (Refer to Section
\ref{sub:Extreme-Chi2}). The more users the system has, the larger
the sum rate is. 

The effect of beamforming can be analyzed according to (\ref{eq:E-n^2-calculation}).
Define $K\triangleq\left|\mathcal{B}\right|$ and $R_{\mathrm{fb}}\triangleq\log_{2}K$.
Assume that $R_{\mathrm{fb}}$ is large so that $\left(l-\gamma\right)\ll\gamma$.
Then approximately, $\mathrm{E}\left[\sum_{j=1}^{l}n_{j}^{2}\right]$
is proportional to $\gamma$. The beamforming feedback function should
maximize $\gamma$ and Assumption F2 is therefore verified. Denote
$l-\gamma_{\sup}=D^{*}\left(K\right)$ the beamforming loss. From
the distortion rate function on the $\mathcal{G}_{L_{T},1}^{\left(l\right)}\left(\mathbb{C}\right)$,
$l-\gamma_{\sup}$ is a exponentially decreasing function of $R_{\mathrm{fb}}/l\left(L_{T}-1\right)$.
We expect that a few feedback bits on beamforming could have large
gain while more feedback bits wouldn't gain much further. 

The assumption T3 about the constant number of on-beams can be validated
as well. Assume that both the number of users $N$ and the feedback
bits on beamforming $R_{\mathrm{fb}}$ are large. Because of the user
choice and beamforming, the quantities $n_{j}^{2}$ $1\leq j\leq l$
are relatively {}``stable'', i.e., the fluctuations of $n_{j}^{2}$'s
are relatively small. It is reasonable to assume constant number of
on-beams for multi-access system. 

The antenna selection can be viewed as a special case of general beamforming
where the beamforming vector is always a column of the identity matrix.
For general beamforming, $\log_{2}\left(\begin{array}{c}
N\\
l\end{array}\right)+R_{\mathrm{fb}}$ feedback bits are needed. For antenna selection, there are $\log_{2}\left(\begin{array}{c}
NL_{T}\\
l\end{array}\right)\approx\log_{2}\left(\begin{array}{c}
N\\
l\end{array}\right)+l\log_{2}L_{T}$ feedback bits needed. Since antenna selection does not assume one
on-beam per on-user (Assumption T2), it is expected that the sum rate
of antenna selection is close to but better than that of general beamforming
with the same $l$ and $R_{\mathrm{fb}}=l\log_{2}L_{T}$. The improvement
is due to the extra freedom the antenna selection has.

\subsection{Simulation }

\begin{figure}[hbt]
\begin{minipage}[c]{0.5\columnwidth}%
\includegraphics[%
  clip,
  scale=0.6]{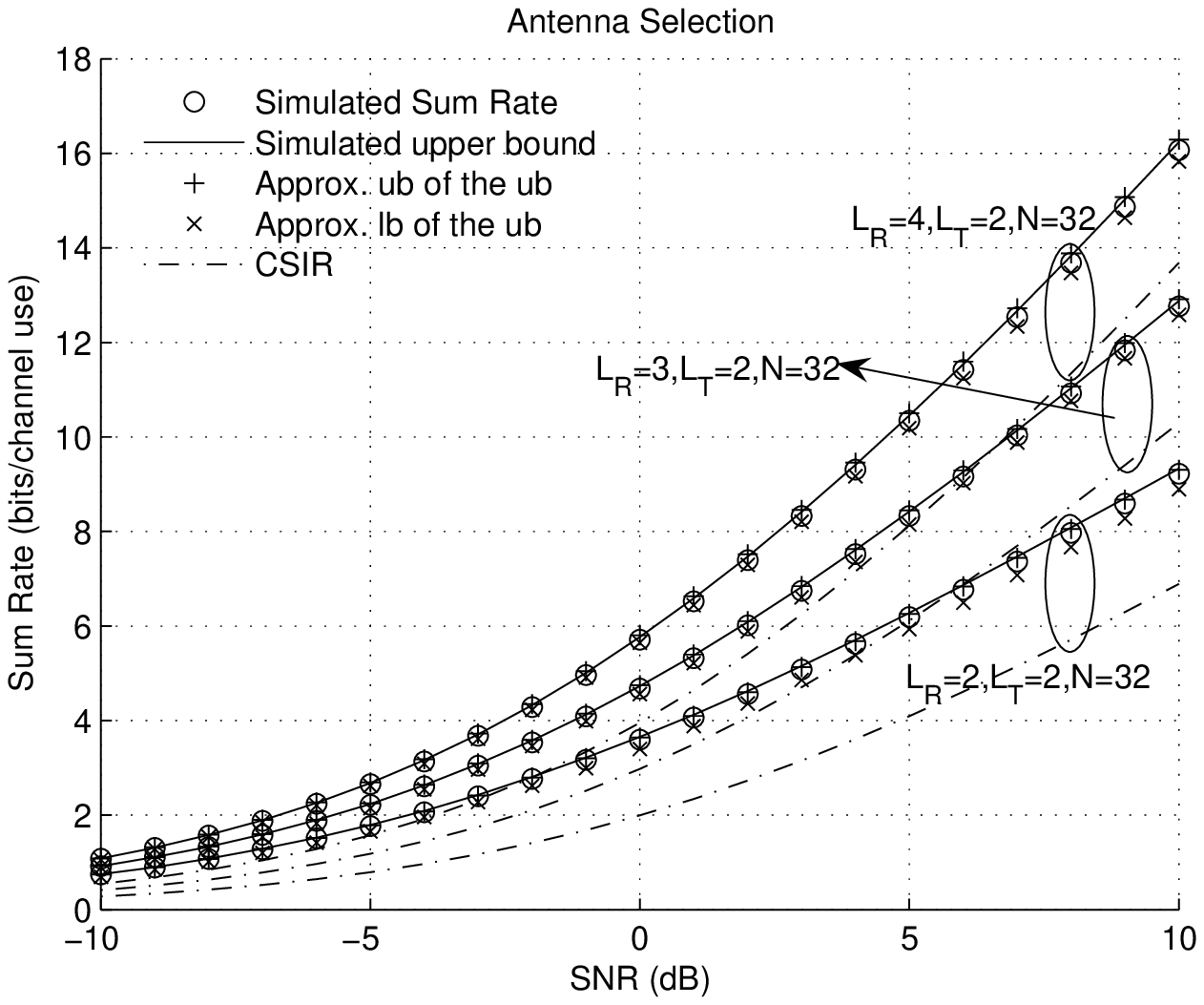}

\caption{\label{cap:Antenna-Selection}Sum Rate for Antenna Selection.}\end{minipage}%
\begin{minipage}[c]{0.5\columnwidth}%
\includegraphics[%
  clip,
  scale=0.6]{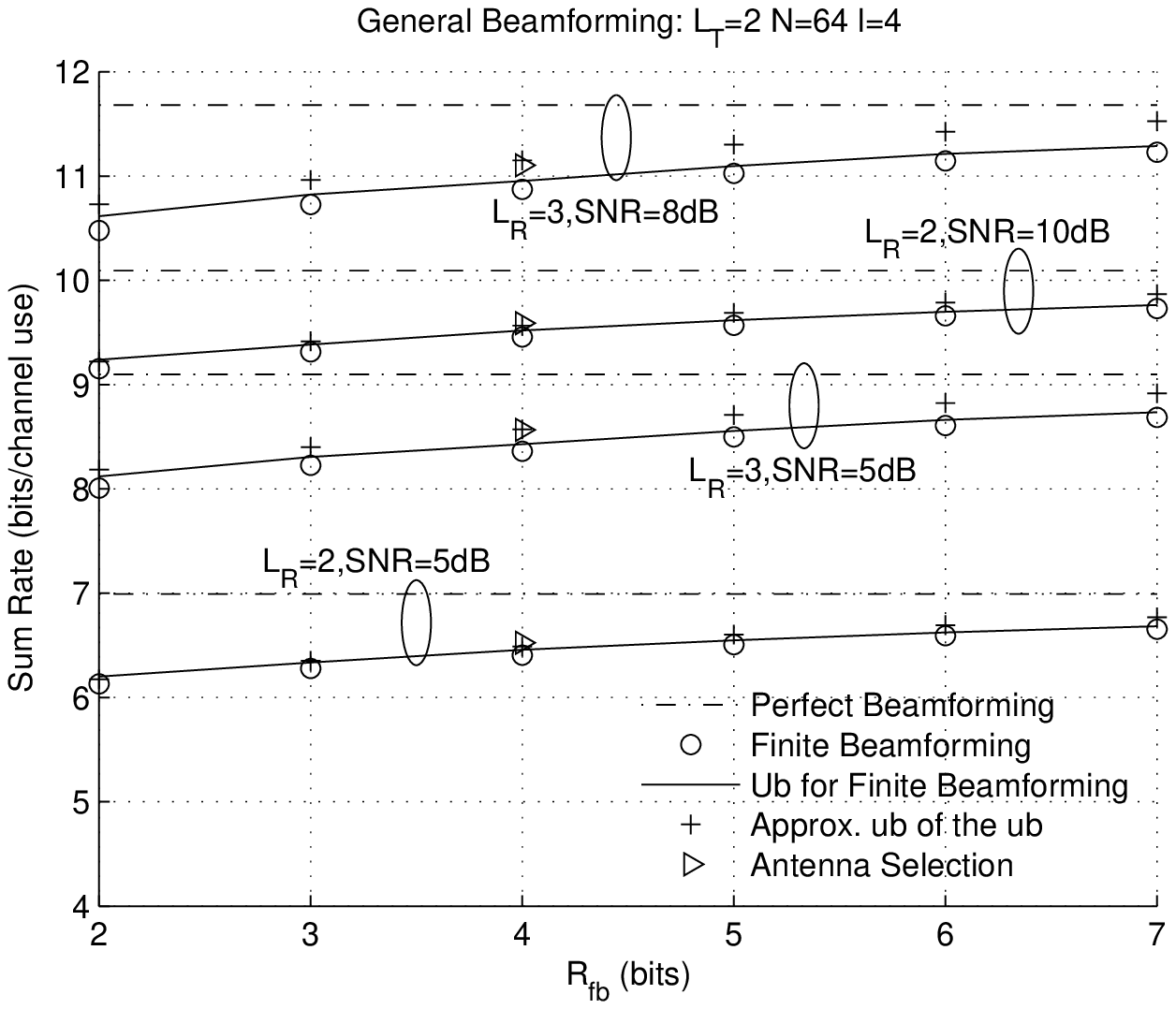}

\caption{\label{cap:Beamforming}Sum Rate for General Beamforming.}\end{minipage}%

\end{figure}

The sum rates of antenna selection and general beamforming are given
in Fig. \ref{cap:Antenna-Selection} and Fig. \ref{cap:Beamforming}
respectively. Simulations show that the upper bound (\ref{eq:ub_beamforming})
(solid lines) is tight. Note that the upper bound (\ref{eq:ub_beamforming})
is of the form $\mathrm{E}\left[\log\left|\mathbf{I}+c\mathbf{\Xi}^{\dagger}\mathbf{\Xi}\right|\right]$.
Theoretical analysis (Theorem \ref{thm:bds_CGMatrix}) gives an upper
bound (plus markers) and a lower bound ('x' markers) on (\ref{eq:ub_beamforming}).
Simulations show that these theoretical approximations are accurate.

Fig \ref{cap:Beamforming} also depicts the gain of beamforming. The
sum rate by finite rate beamforming feedback (circles) is compared
to that of perfect beamforming (dash-dot lines). Simulation shows
that with several feedback bits on beamforming, the corresponding
sum rate is close to that of perfect beamforming. As a special case
of general beamforming, antenna selection is shown to be similar to
but better than general beamforming with the same $l$ and $R_{\mathrm{fb}}=l\log_{2}L_{T}$.

\section{Conclusion}

This paper proposes a strategy where users are controlled jointly.
The effect of user choice is analyzed by extreme order statistics
and the effect of beamforming is quantified by the distortion rate
function in the composite Grassmann manifold. By characterizing the
distortion rate function on the composite Grassmann manifold and calculating
the logdet function of a random composite Grassmann matrix, a good
sum rate approximation is derived. 

\bibliographystyle{IEEEtran}
\bibliography{Bib/_Heath,Bib/_Liu_Dai,Bib/_love,Bib/_Rao,Bib/_Tse,Bib/FeedbackMIMO_append,Bib/RandomMatrix}

\end{document}